\documentclass[aps,prl,reprint,groupedaddress]{revtex4-1}

\usepackage{graphicx}
\usepackage[colorlinks=true,linkcolor=blue,citecolor=blue]{hyperref}
\usepackage{amsmath,amssymb}
\usepackage{mathrsfs,amssymb}

\begin{document}

\title{Superresolved optical imaging through higher-order spatial frequency harmonic generation without beating the diffraction limit of light}
\author{Zhixiang Li}
\affiliation{The MOE Key Laboratory of Weak Light Nonlinear Photonics, School of Physics and TEDA Applied Physics Institute, Nankai University, Tianjin 300457, China}

\author{Jianji Liu}
\affiliation{The MOE Key Laboratory of Weak Light Nonlinear Photonics, School of Physics and TEDA Applied Physics Institute, Nankai University, Tianjin 300457, China}

\author{Guoquan Zhang}
\email[Corresponding author:]{zhanggq@nankai.edu.cn}
\affiliation{The MOE Key Laboratory of Weak Light Nonlinear Photonics, School of Physics and TEDA Applied Physics Institute, Nankai University, Tianjin 300457, China}

\date{\today}

\begin{abstract}
We proposed a method to achieve superresolved optical imaging without beating the diffraction limit of light. This is achieved by magnifying the ideal optical image of the object through higher-order spatial frequency generation while keeping the size of the effective point spread function of the optical imaging system unchanged.
A proof-of-principle experiment was demonstrated in a modified $4f$-imaging system, where the spatial frequency of a two-line source was doubled or tripled on the confocal Fourier plane of the $4f$-imaging system through a light pulse storage and retrieval process based on the electromagnetically induced transparency effect in a Pr$^{3+}$:$\rm Y_2 SiO_5$ crystal, and an originally unresolvable image of the two line sources in the conventional $4f$-imaging system became resolvable with the spatial frequency doubling or tripling. Our results offer an original way towards improving optical imaging resolution without beating the diffraction limit of light, which is totally different from the existing superresolution methods to overcome the diffraction limit.
\end{abstract}


\maketitle

Optical imaging technique plays an important role in various areas such as optics, astronomy and biology. In general, the image $U_i(x_i,y_i)$ of the object can be viewed as the result of a convolution of the ideal image predicted by geometrical optics $U_g(x_i,y_i)$ and the point spread function (PSF) $h(x_i,y_i)$ of the optical imaging system~\cite{Goodman1996}
\begin{equation}\label{convolution}
U_i(x_i,y_i)=\int h(x_i-x,y_i-y)U_g(x,y)dxdy \, .
\end{equation}

\noindent The spatial resolution of the optical imaging system, due to the wave property of light, is characterized by PSF. According to the Rayleigh criteria, the spatial resolution is limited to be $0.61\lambda/\rm NA$ for a diffraction-limited microscopy, where $\lambda$ is the wavelength of light in vacuum and $\rm NA$ is the numerical aperture of the objective lens.

Various methods, both near-field and far-field ones, were developed to overcome the diffraction limit in order to distinguish tiny structures beyond the diffraction limit. Typical near-filed techniques such as scanning near-field optical microscope~\cite{Betzig1986} achieve superresolved imaging by gathering higher spatial frequency in the near field, while typical far-field fluorescence microscopies overcome the diffraction limit by temporally activating or spatially modulating neighboring fluorophores in order to precisely locate the imaging point~\cite{Huang2010,Betzig2015}. For example, in the photo-activated localization microscopy (PALM)~\cite{Betzig2006} and stochastic optical reconstruction microscopy (STORM)~\cite{Rust2006}, individual molecule is stochastically activated within the diffraction-limited region at different times, therefore, can be located separately to overcome the diffraction limit of light.
In stimulated emission depletion (STED)~\cite{Hell1994,Rittweger2009}, the excitation pulse excites the fluorophores to their fluorescent state and the ring-shaped STED pulse de-excites the fluorophores by means of stimulated emission, leading to a reduced PSF.
Structured illumination microscopy~\cite{Gustafsson2000} and saturated structured-illumination microscopy~\cite{Gustafsson2005} achieve super-resolution by illuminating a sample with patterned light and the normally inaccessible high-resolution information can be gathered by measuring the fringes in the Moir$\acute{e}$ pattern.
Note that the key of the above far-field superresolved microscopies is to reduce the size of effective PSF of the imaging system~\cite{Huang2010,Betzig2015}.

In this Letters, we proposed a different method to discern a tiny structure with a size scale smaller than the diffraction limit of light. Instead of trying to narrow the size of effective PSF, we magnify the ideal image $U_g$ through a high-order spatial frequency generation process while keeping the size of effective PSF unchanged. This makes the separation distance between two neighboring points increase but without size expansion of diffraction spots, therefore, leading to a superresolved imaging.

\begin{figure}[t]
\centering\includegraphics[width=7.5cm]{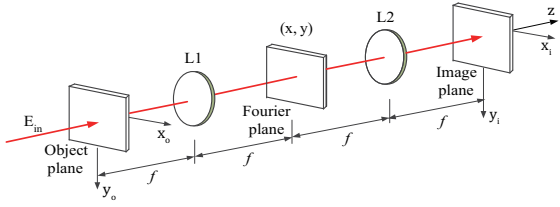}
\caption{The schematic diagram of a conventional $4f$-imaging system, where L1 and L2 are two lens with a focal length $f$. }
\label{4fscheme}
\end{figure}

To illustrate the superresolving principle but without loss of generality, let us take a $4f$-imaging system (see Fig.~\ref{4fscheme}) as an example. In an ideal optical imaging system without considering the diffraction of light, the field amplitude in the confocal Fourier plane is expressed as
\begin{equation}\label{imaging_Eq}
U_F(x,y)=\frac{e^{i2kf}}{i\lambda f}\int U_o(x_o,y_o)e^{-i(kx_o x+ky_o y)/f} dx_o dy_o \, ,
\end{equation}
\noindent where $f$ is the focal length of lens, $k=2\pi/\lambda$ is the wave number of light, $x_o$, $y_o$ are the transverse coordinates in the object plane, and $x$, $y$ are the transverse coordinates in the Fourier plane with ($kx/f$, $ky/f$) being the spatial frequency of object, respectively. The field amplitude in the imaging plane can then be given as
\begin{eqnarray}
U_g(x_i,y_i)&=&\frac{e^{i2kf}}{i\lambda f}\int U_F(x,y)e^{-i(kxx_i+kyy_i)/f} dxdy  \nonumber \\
  &=&-e^{i4kf} U_o(-x_i,-y_i)
\end{eqnarray}
\noindent Note that the negative sign in ($-x_i$, $-y_i$) indicates an inversed image with respect to the object.

Now, if one introduces $n$th-order harmonics for all spatial frequencies of the object, i.e., replacing $U_F(x,y)$ with $U_F(nx,ny)$ in the Fourier plane, one gets
\begin{equation}
U_g(x_i, y_i)=-\frac{e^{{i4kf}}}{n^2} U_o\left(-\frac{x_i}{n}, -\frac{y_i}{n}\right) \,  ,
\end{equation}
\noindent indicating that the image is magnified by $n$ times in space. It is evident that, if one keeps the size of effective PSF of the optical imaging system unchanged during the introduction of high-order harmonics of the object in the Fourier plane, the spatial resolution of the image, which can be characterized by finesse $F$ defined as the ratio between the separation distance of two imaging peaks $d_i$ and the width of the imaging peak $w_i$~\cite{Li2008}, will be improved by $n$ times according to Eq.~(\ref{convolution}).

It seems unpractical to magnify only the ideal image $U_g$ without the simultaneous size expansion of PSF in a conventional optical system. But this is indeed possible, as we will demonstrate in the following, in a modified $4f$-imaging system, in which a light pulse storage and retrieval process based on electromagnetically induced transparency (EIT) effect~\cite{Harris1997,Fleischhauer2005,Lukin2001,Fleischhauer2000} is performed in the confocal Fourier plane of the $4f$-imaging system. In a coherent atomic ensemble driven by a strong coupling beam, the absorption of a probe beam, even resonantly applied on an atomic transition, can be reduced or even eliminated under the EIT condition due to the quantum destructive interference between two resonant atomic transition amplitudes. According to the Kramers-Kronig relationship, the EIT effect will induce a strong spectral dispersion near the two-photon resonant condition, which was used to slowdown and stop light pulses propagating in the EIT media. Storage and retrieval of light pulses carrying optical information~\cite{Liu2001,Pugatch2007,Cho2010,Zhai2013} or encoded with images~\cite{Camacho2007,Vudyasetu2008,Shuker2008} were demonstrated based on the EIT effect.

\begin{figure}[t!]
\centering\includegraphics[width=8.5cm]{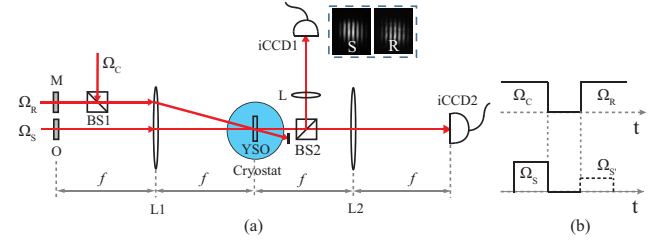}
\caption{(a) A modified $4f$-imaging system with EIT-based light pulse storage and retrieval in the confocal Fourier plane. Here $\Omega_S$, $\Omega_C$, $\Omega_R$ and $\Omega_{S'}$ are the signal, the coupling, the readout and the retrieved signal beams, respectively. O and M are the object and the $\pi$-phase-shifted $n$-slit mask, BS1 and BS2 are the beam splitter, and L1, L2 and L are lens, respectively.  The inset shows the intensity distribution of the signal beam $\Omega_S$ and the readout beam $\Omega_R$ in the confocal Fourier plane. (b) the time sequence of $\Omega_S$, $\Omega_C$, $\Omega_R$ and $\Omega_{S'}$ during the light pulse storage and retrieval process.}
\label{setup}
\end{figure}

For simplicity but without loss of generality, we consider a light pulse storage and retrieval process  based on EIT effect in an atomic ensemble with a typical three-level $\Lambda$-type energy-level configuration.
The experimental setup is shown in Fig.~\ref{setup}.  A praseodymium ion doped $\rm Y_2 SiO_5$ crystal ($\rm Pr^{3+}$:YSO, 0.05$\%$) was employed as the EIT media, and two lower levels and one upper level of the hyperfine transition $^3H_4\leftrightarrow{}^1D_2$ ($\sim$ 605.78 nm) of $\rm Pr^{3+}$ ions were chosen to form the three-level $\Lambda$-type energy level configuration~\cite{Tu2009,Zhai2011}. The crystal was kept at 3.6 K in a cryostat and was placed in the confocal Fourier plane of the $4f$-imaging system with a focal length of lens $f=30$ cm. Three beams from a Coherent 899 dye ring laser operating at $\sim 605.78$ nm served as the probe beam $\Omega_S$, the coupling beam $\Omega_C$ and the readout beam $\Omega_R$ with their respective power of 16 mW, 30 mW  and 30 mW. These beams could be temporally modulated in intensity and shifted in frequency independently by the corresponding acousto-optic modulators (not shown here), and they were polarized linearly and in parallel in such a way to guarantee an efficient EIT effect and storage efficiency~\cite{Zhai2012}. A 5-$\rm \mu s$ probe pulse, after transmitted through an object at the front focal plane of lens L1, became the object beam and interacted with the coupling beam in the $\rm Pr^{3+}$:YSO crystal under the EIT condition in the confocal Fourier plane, and the object field carrying the Fourier spatial frequency of the object was stored in $\rm Pr^{3+}$:YSO crystal by adiabatically switching off the coupling field. After a certain time delay ($\sim$ 3 $\rm \mu s$ in our experiments, less than the storage time of light pulse), a new field $\Omega_{S^{^\prime}}$ was obtained by launching the readout beam $\Omega_R$ under the phase-matching condition. The time sequence of the interacting beams is shown in Fig.~\ref{setup}(b), and  detailed procedure to store and retrieve light pulse can be found in Refs.~\cite{Tu2009,Zhai2011}. The light intensity distribution on the Fourier plane was imaged through lens L and recorded by an intensified charge-coupled device (iCCD1) camera, and the image of the object through the $4f$-imaging system was detected by another iCCD2.

Assuming that the light field amplitude on the confocal Fourier plane is $E_{j}(x,y)$ (j=$S$, $C$, $R$, $S^\prime$) for the object, the coupling, the readout and the retrieved signal beams, respectively. After a light-pulse storage and retrieval process under the phase-matching condition, the field amplitude of the retrieved signal beam $E_{S^\prime}(x,y)$, that is also $U_F(x,y)$, can be written as~\cite{Zhai2012,Zhai2011}:
\begin{equation}\label{EqEIT}
U_F(x,y)=E_{S^\prime}(x,y)\propto E_{S}(x,y)E^*_{C}(x,y)E_{R}(x,y) \,  ,
\end{equation}
\noindent
where the superscript $'*'$ denotes the complex conjugate.

As a proof-of-principle demonstration and to show the superresolving capability, we consider the one-dimensional case, where the imaging of two line sources separated by a distance $d$ and placed in the front focal plane of lens L1 is considered. Due to the light diffraction, the images of the line sources will no longer be a geometric line but with a finite spatial linewidth $w_i$, which is determined and can be characterized by the full-width at half maximum (FWHM) of the effective PSF of the imaging system. The separation distance between two imaging line peaks of two line sources is $d_i$ and the finesse of the imaging system is then written as $F=d_i/w_i$.

Ideally, the light field amplitude in the confocal Fourier plane is proportional to $\cos(\pi dx/\lambda f)$.  In the experiment, a double-slit mask with a slit-width $a$ and a separation distance $d$ between two slits was employed to mimic the two line sources. Note that the introduction of finite slit width $a$, which acts as an optical aperture as far as line sources are considered, will modify the effective PSF of the $4f$-imaging system.  In this case, the field amplitude $E_S(x)$ in the confocal Fourier plane can be written as
$
E_S(x) \propto {\rm{\cos}}\left(\frac{{\pi dx}}{\lambda f}\right){\rm{sinc}}\left(\frac{{\pi ax}}{\lambda f}\right)
$,
where the interference term $\cos(\pi dx/\lambda f)$  contains the spatial frequency to form the ideal image of two line sources, while the term ${\rm{sinc}}(\pi a x/\lambda f)$ describes the diffraction effect of a single slit. For the ideal two  line sources, the effective PSF $h_{eff}(x_i)$ is the convolution of a rectangular function ${\rm rect}(x_i/a)$ and the PSF of a conventional $4f$-imaging system $h(x_i)=h_0 {\rm sinc}(\pi A x_i/\lambda f)$, where $h_0$ is a normalization factor, and $A$ is the size of effective optical aperture of the imaging system, i.e., the width of the transmission slit on the sample holder in the cryostat here. The information carried by $E_S(x)$ was then stored in Pr$^{3+}$:YSO crystal based on EIT effect by adiabatically switching off the Gaussian coupling beam, which can be approximately characterized by a constant field amplitude $E_C$ when its spot size is much larger than those of object and readout beams.

To magnify the distance between two imaging line peaks $d_i$ while keeping the diffraction effect unchanged, one can generate high-order spatial frequency harmonics in the Fourier plane, for example, the second-order harmonics, but without size expansion of the effective PSF. This can be realized by employing a readout beam with a field amplitude proportional to $\sin(\pi dx/\lambda f)$, which can be generated by interfering two coherent beams. In the experiment, we let the readout beam $\Omega_R$ transmit through another double-slit mask with the same silt width $a$ and separation distance $d$ as the mimicking object mask, which was placed also on the front focal plane of lens L1, as shown in Fig.~\ref{setup}. By introducing a $\pi$-phase-shift on one slit of the double slit, one could produce a diffraction light field $E_R(x)\propto \sin(\pi dx/\lambda f) {\rm sinc}(\pi ax /\lambda f)$ in the Fourier plane, which was phase-shifted by $\pi/2$ with respect to $E_S(x)$.  Note that there is an additional diffraction term ${\rm sinc}(\pi ax/\lambda f)$ associated with $E_R(x)$, which will modify further the effective PSF of the imaging system. Fortunately, as we will show below, the introduction of this additional diffraction term will not lead to a size expansion of the effective PSF as compared to that of the conventional $4f$-imaging system without EIT-based light pulse storage and retrieval process.

According to Eq.~(\ref{EqEIT}), the retrieved field amplitude can then be expressed as
\begin{equation}\label{8}
 {E_{{\rm{S^\prime}}}}(x) \propto \sin \left(\frac{{\pi(2d)x}}{\lambda f}\right){\rm{sin}}{{\rm{c}}^2}\left(\frac{{\pi{ax}}}{\lambda{f}}\right)\ \, .
\end{equation}
Then a Fourier transform through lens L2 produces an image of two line sources with a separation distance $d_i=2d$, which is twice as that in a conventional $4f$-imaging system. Note that, according to Eq.~(\ref{8}), the effective PSF $h_{eit}(x_i)$ in this case is a convolution of a triangle function $tri(x_i)$ and $h(x_i)$, where $ tri(x_i)=1-|x_i/a|$ when $|x_i|<a$, while $tri(x_i)=0$ if $|x_i|\geq a$. One can confirm theoretically that the imaging linewidth $w_i$ is the same for $h_{eff}(x_i)$  and $h_{eit}(x_i)$.
Therefore, as compared to the case with a conventional $4f$-imaging system, a finesse improvement $\sim$2 can be obtained theoretically with our modified $4f$-imaging system.

\begin{figure}[b]
\centering\includegraphics[width=8.4cm]{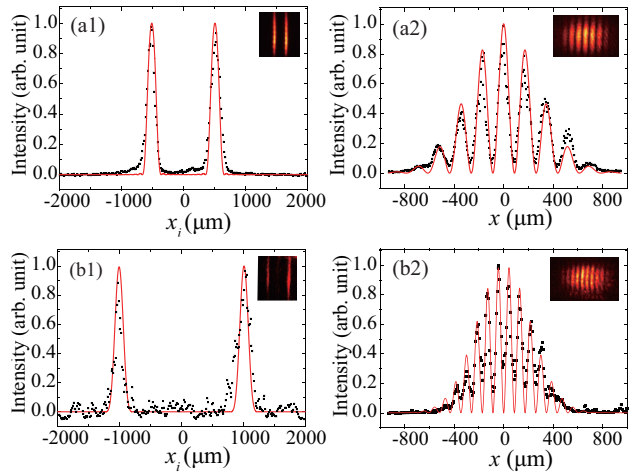}
\caption{The transverse intensity profiles of images of two line sources mimicked by a double-slit  with $d=1000$ $\rm \mu m$ and $a=200$ $\rm \mu m$ in the conventional (a1) and the modified (b1) $4f$-imaging systems, respectively. The insets are the corresponding images of two line sources recorded by iCCD2. (a2) and (b2) are the transverse spatial spectra on the confocal Fourier plane of the conventional and the modified $4f$-imaging systems, respectively. The insets show the corresponding intensity distribution recorded by iCCD1 through L. The red solid curves are theoretical simulation with experimental parameters without any fitting parameter. }
\label{exp_second}
\end{figure}

Figures~\ref{exp_second}(a1) and~\ref{exp_second}(b1) show transverse intensity profiles of images of two line sources mimicked by a double-slit with $d=1000$ ${\rm \mu m}$ and $a=200$ ${\rm\mu m}$ in the conventional and the modified $4f$-imaging systems, respectively, and the insets show the corresponding images recorded by iCCD. One sees that the separation distance between two imaging line peaks $d_i$ in Fig.~\ref{exp_second}(b1) is twice as that in Fig.~\ref{exp_second}(a1), while the imaging linewidth $w_i$ was measured to be 148 $\rm\mu m$ in both cases, indicating a finesse improvement of $\sim 2$, which is in good agreement with the theoretical prediction. Figures~\ref{exp_second}(a2) and~\ref{exp_second}(b2) are the transverse spatial spectra on the confocal Fourier plane in the conventional and modified $4f$-imaging systems, respectively. The interference period in Fig.~\ref{exp_second}(b2) was measured to be 91 $\rm\mu m$, which is half of that in Fig.~\ref{exp_second}(a2). This verified that the finesse improvement in our modified $4f$-imaging system is originated from the spatial frequency doubling effect achieved through the EIT-based light-pulse storage and retrieval process. The red solid curves in Fig.~\ref{exp_second} are simulated with experimental parameters $d=1000$ $\rm \mu m$, $a=200$ $\rm \mu m$ and $A=2.4$ mm, and no fitting parameter was used in the simulation. Good agreement is seen between theoretical prediction and experimental measurements.

The finesse in the modified $4f$-imaging system can be further increased when $n$th-order ($n>2$) spatial frequency harmonics is introduced. In general, one can use a $n$-slit mask with the same slit width and separation distance as the double-slit used to mimic the two line sources, and sets the light fields transmitting through two neighboring slits of $n$-slit mask out of phase, i.e., introduces a $\pi$-phase-shift between two nearest slits of $n$-slit mask. We call such kind of mask as $\pi$-phase-shifted $n$-slit mask. Similar to the light pulse storage and retrieval procedure to generate the 2nd-order harmonics described above, by replacing the $\pi$-phase-shifted double-slit mask with the $\pi$-phase-shifted $n$-slit mask in the optical path of readout beam $\Omega_R$, one can obtain a retrieved signal field as
\begin{equation}\label{even}
E_{S\prime}(x)\propto\sin\left(\frac{\pi ndx}{\lambda f}\right){\rm sinc}^2 \left(\frac{\pi a x}{\lambda f}\right)
\end{equation}
when $n$ is even integer, or
\begin{equation}\label{odd}
E_{S\prime}(x)\propto\cos\left(\frac{\pi ndx}{\lambda f}\right){\rm sinc}^2 \left(\frac{\pi a x}{\lambda f}\right)
\end{equation}
when $n$ is odd integer. Then the separation distance between the two imaging line peaks $d_i$ will be $nd$, while the imaging linewidth $w_i$ will be the same as compared to that of the conventional $4f$-imaging system.

\begin{figure}[t]
\centering\includegraphics[width=8.4cm]{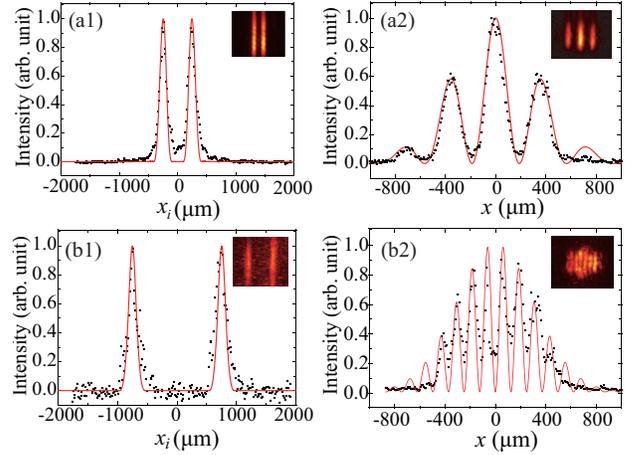}
\caption{The transverse intensity profiles of images of two line sources mimicked by a double-slit with $d=500$ $\rm \mu m$ and $a=200$ $\rm \mu m$ in the conventional (a1) and the modified (b1) $4f$-imaging systems, respectively. The insets are the corresponding images of the same two-line source recorded with iCCD2. (a2) and (b2) are the transverse spatial spectra on the confocal Fourier plane of the conventional and the modified $4f$-imaging systems, respectively. The insets show the corresponding intensity distribution recorded by iCCD1 through L. The red solid curves are simulated with experimental parameters $d=500$ $\rm \mu m$, $a=200$ $\rm \mu m$ and $A=2.4$ mm.}
\label{thirdharmonics}
\end{figure}

Figure~\ref{thirdharmonics} shows the results with third harmonic generation, where the two line sources were mimicked by a
double-slit with $d=500$ $\rm\mu m$ and $a=200$ $\rm\mu m$, and a $\pi$-phase-shifted three-slit mask with the same $d=500$ $\rm\mu m$ and $a=200$ $\rm\mu m$  was inserted in the optical path of readout beam $\Omega_R$. One sees from Figs.~\ref{thirdharmonics}(a1) and~\ref{thirdharmonics}(b1) that the separation distance between two imaging line peaks $d_i$  in our modified $4f$-imaging system was tripled. The imaging linewidth $w_i$ was measured to be 148 $\rm \mu m$, which, again, is the same as that in the conventional $4f$-imaging system. This leads to a finesse increase of $\sim 3$ in comparison to the conventional $4f$-imaging system. By comparing the periodicity of the spatial spectra in the confocal Fourier plane of the conventional and modified $4f$-imaging systems (see Figs.~\ref{thirdharmonics}(a2) and~\ref{thirdharmonics}(b2)), we verified that the observed finesse increase is originated from the spatial frequency tripling effect achieved through the EIT-based storage and retrieval process. It is evident that the finesse of the modified $4f$-imaging system will increase further on through higher-order spatial frequency generation.

\begin{figure}[t]
\centering\includegraphics[width=6.0cm]{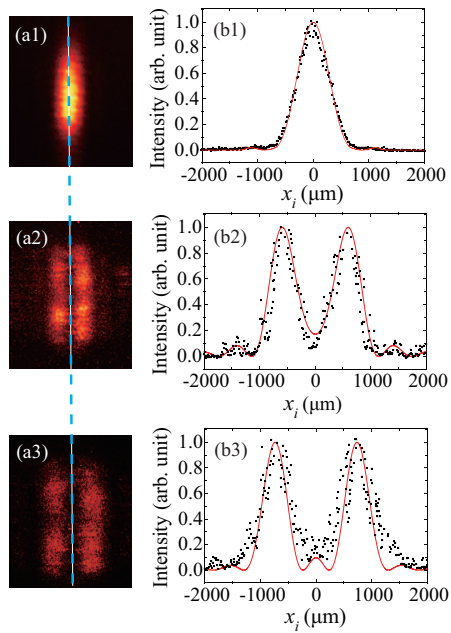}
\caption{Images and the corresponding transverse intensity profiles of the two line sources mimicked by a double-slit with $d=500$ $\rm \mu m$ and $a=200$ $\rm \mu m$ in the conventional $4f$-imaging system ((a1) and (b1)), the modified $4f$-imaging system with spatial frequency doubling ((a2) and (b2)) and tripling ((a3) and (b3)), respectively. The red solid curves are simulated with $d=500$ $\rm \mu m$, $a=200$ $\rm \mu m$ and $A=310$ $\rm \mu m$.}
\label{superresolving}
\end{figure}

Finally, we demonstrated the capability for superresolved imaging even without beating the diffraction limit of the conventional $4f$-imaging system. The two line sources were again mimicked by a double-slit with $d=500$ $\rm\mu m$ and $a=200$ $\rm\mu m$. By narrowing the transmission slit width $A$ on the sample holder in cryostat down to 310 $\rm \mu m$, the images of two line sources were overlapped and unresolvable in the conventional $4f$-imaging system, as shown in Figs.~\ref{superresolving}(a1) and~\ref{superresolving}(b1). Then, we kept the aperture width $A$ fixed and doubled or tripled the spatial frequency in the confocal Fourier plane of the modified $4f$-imaging system through the light-pulse storage and retrieval process as described above. The images of the two line sources became resolved, as shown in Figs.~\ref{superresolving}(a2)  and \ref{superresolving}(b2) and Figs.~\ref{superresolving}(a3) and \ref{superresolving}(b3), respectively, for the spatial frequency doubling and tripling cases.  Note that the imaging linewidth $w_i$ of each line source is the same for all three cases in Fig.~\ref{superresolving}, showing the superresolving capability of this technique even without beating the diffraction limit of the conventional $4f$-imaging system.

We note that optical imaging beyond the diffraction limit~\cite{Li2008,Verma2013} and subwavelength selective localization of atomic excitation~\cite{Agarwal2006,Yavuz2007,Gorshkov2008,Priote2011,Miles2013} have been reported under the EIT condition, where the key idea is the sensitivity of dark state of EIT to the coupling beam intensity, leading to nonlinear dependence of the absorption or refraction index and the population of atomic level on the spatial intensity distribution of the coupling beam. Therefore, the mechanism to achieve subwavelength resolution in these techniques is totally different from ours, especially with respect to the diffraction limit of light. We emphasize that the superresolved imaging is achieved without beating the diffraction limit of light in our case.

In conclusion, we have achieved superresolved imaging through higher-order spatial harmonic generation based on light pulse storage and retrieval process under the EIT condition even without beating the diffraction limit of light. Moreover, as compared to STORM, PALM and STED, fluorescence labeling is not necessary in our case.

The authors thank Dr. P. Hong, Z. Zhai and L. Huang for helpful discussions.
This project is supported by the 973 program (2013CB328702), the NSFC (11174153 and 61475077), the 111 project (B07013) and the PCSIRT (IRT-13R29).

\end{document}